\documentclass[aps,preprint]{revtex4}%
\usepackage{amsfonts}
\usepackage{amsmath}
\usepackage{amssymb}
\usepackage{graphicx}%
\begin{document}
\title{Magnetoplasmon spectrum of a weakly modulated two-dimensional electron gas system}
\author{M. Tahir$^{1,\ast}$, K. Sabeeh$^{2,\dag}$ and A. MacKinnon$^{1}$}
\affiliation{$^{1}$Department of Physics, Blackett Laboratory, Imperial College London,
South Kensington Campus, London SW7 2AZ, United Kingdom.}
\affiliation{$^{2}$Department of Physics, Quaid-i-Azam University Islamabad 45320,
Pakistan. }

\pacs{72.80.Rj, 71.45.Gm, 73.21.-b}

\begin{abstract}
The magnetoplasmon spectrum of a magnetically modulated two-dimensional
electron gas (MM2DEG) is investigated. We derive the inter and intra Landau
band magnetoplasmon spectrum within the self consistent field approach. The
derivation is performed at zero temperature as well as at finite temperature.
Results are presented for the inter and intra Landau band magnetoplasmon
spectrum as a function of the inverse magnetic field. As a result of magnetic
modulation, magnetic Weiss oscillations are found to occur in the
magnetoplasmon spectrum. Furthermore, our finite temperature theory
facilitates analysis of effects of temperature on the magnetoplasmon spectrum.
The results are compared with those obtained for an electrically modulated
2DEG system. In addition, we derive and discuss the effects of simultaneous
electric and magnetic modulations on the magnetoplasmon spectrum of 2DEG when
the modulations are in phase as well as when they are out of phase. Magnetic
oscillations are affected by the relative phase of the two modulations and
position of the oscillations depends on the relative strength of the two
modulations in the former case while we find complete suppression of Weiss
oscillations for particular relative strength of the modulations in the latter case.

\end{abstract}
\date[Date text]{date}
\received[Received text]{date}

\revised[Revised text]{date}

\accepted[Accepted text]{date}

\published[Published text]{date}

\maketitle

\section{Magnetoplasmon spectrum of magnetically modulated two-dimensional
electron gas (MM2DEG)}

In the past two decades, remarkable progress has been made in epitaxial
crystal growth techniques which have made possible the fabrication of novel
semiconductor heterostructures. These modern microstructuring techniques can
be used to laterally confine quasi-two-dimensional electron gas (2DEG) in
e.g.,GaAs/AlGaAs heterostructure on a submicrometer scale. Furthermore,
magnetic modulation of these systems can be realized by depositing an array of
ferromagnetic strips on top of the heterostructure or by superconducting
layers beneath the substrate. These magnetically modulated 2DEG systems
realized in semiconductor heterostructures have attracted a lot of attention
in the past and continue to do so today due to their novel
properties\cite{1,2,3,4,5,6,7,8,9,
10,11,12,13,14,15,16,17,18,19,20,21,22,23,24,25,26,27,28}. Primarily, it is
due to the introduction of another length scale in the system, the period of
modulation, as a result new phenomena arise due to commensurability of the
period of modulation and the other characteristic length scale of the system,
cyclotron diameter at the Fermi level. Resistivity measurements on these
systems are found to exhibit commensurability oscillations (magnetic Weiss
oscillations) since the magnetic modulation broadens the Landau levels into
minibands whose width oscillates as a function of the magnetic field. These
oscillations are periodic as a function of the inverse magnetic field with a
larger period than that of the Shubnikov-de Hass (SdH) oscillations. The
period of Weiss oscillations depends on both the modulation period and the
square root of the number density of the MM2DEG, in contrast to the linear
dependence on the number density of the SdH oscillations. Moreover, the
amplitude of Weiss oscillations is weakly affected by temperature as compared
to SdH oscillations.

We investigate the effects of magnetic modulation on the collective
excitations (magnetoplasmons) of a 2DEG. Plasmons are a very general
phenomenon and have been studied extensively in a wide variety of systems
including ionized gases, simple metals and semiconductor 2DEG systems. In a
2DEG, these collective excitations are induced by the electron-electron
interactions. These collective excitations, plasmons, are among the most
important electronic properties of a system. In the presence of an external
magnetic field, these collective excitations are known as magnetoplasmons.
Magnetic oscillations of the plasmon frequency occur in a magnetic field.
Single particle magneto-oscillatory phenomena such as the Shubnikov-de Haas
and de Haas-van Alphen effects have provided very important probes of the
electronic structure of solids. Their collective analog yields important
insights into collective phenomena. For this reason we study the effects of
magnetic modulation on the collective excitations (magnetoplasmons) in a 2DEG.

We present the magnetoplasmon spectrum of a MM2DEG in the presence of a
perpendicular magnetic field using the self consistent field approach. Both
the inter and intra Landau band magnetoplasmons are determined. Inter Landau
band magnetoplasmons arise due to electronic transitions between different
Landau bands whereas intra Landau band magnetoplasmons are a result of
electronic transitions in a single Landau band. We evaluate the dynamic,
nonlocal density-density response function to obtain these results. The
collective excitations in a MM2DEG system have been investigated in the
past\cite{12,16,20,21,24}. All of these studies primarily investigate the
inter Landau band magnetoplasmon mode. Reference\cite{16} is an experimental
study where the inter Landau band magnetoplasmons are investigated by infrared
optical measurements whereas \cite{12,20,21,24} are theoretical studies. In
\cite{12} only the primary inter Landau band magnetoplasmon was considered.
Far-infrared absorption of MM2DEG was theoretically investigated in \cite{20}
to explore plasma oscillations in the system. Similarly plasma oscillations
were also theoretically investigated in \cite{21,24} taking into account
primarily the inter Landau band magnetoplasmons. Since the existence of an
intra-Landau band magnetoplasmon is a result of the finite Landau bandwidth
caused by modulation, the study of collective excitations and in particular
modulation induced effects in this system requires taking into account the
intra Landau band magnetoplasmons. To overcome this shortcoming in previous
work mentioned above, we determine both the intra and inter-Landau band
magnetoplasmon spectrum in this work. In addition, our finite temperature
theory facilitates the analysis of effects of temperature on the Weiss \& SdH
oscillations in the magnetoplasmon spectrum of MM2DEG. Furthermore, we carry
out a detailed comparison of phase and amplitude of magnetic Weiss
oscillations and the electric Weiss oscillations in a 2DEG. We also present
the effects of simultaneous electric and magnetic modulations on the
magnetoplasmon spectrum of a 2DEG. To the best of our knowledge, the complete
study of both the inter and intra Landau band magnetoplasmon spectrum in this
system and comparison of these with the results for the electrically modulated
two-dimensional electron gas system (EM2DEG) \cite{29,30,31,32} has not been
carried out so far. To this end, we have undertaken the present study.

In section II, we present the formulation of the problem. Section III contains
the magnetoplasmon spectrum of a MM2DEG and comparison with an EM2DEG at zero
temperature ($T=0$) whereas in section IV we discuss the temperature
dependence of magnetoplasmon spectrum of a MM2DEG and its comparison with an
EM2DEG including the asymptotic description. In the following two sections,
magnetoplasmon spectrum in the presence of simultaneous electric and magnetic
modulations is presented. We discuss the effects of in-phase modulations
(electric and magnetic) in section V while out of phase is discussed in
section VI. Concluding remarks are made in section VII.

\section{Formulation}

The system that we are considering is a 2DEG in the presence of a
perpendicular magnetic field that is modulated weakly and periodically along
one direction. We take the magnetic field $B$ to be perpendicular to the $x-y$
plane in which electrons with unmodulated areal density $n_{D}$, effective
mass $m^{\ast}$ and charge $-e$ are confined. We employ the Landau gauge and
write the vector potential as $A=(0,Bx+(B_{0}/K)\sin Kx,0)$, where $K$ is
$2\pi/a,$ $a$ is the period of modulation and $B_{0}$ is the magnetic
modulation strength such that $B_{0}\ll B.$ The Hamiltonian in the Landau
gauge is\cite{5,8,26}
\begin{equation}
H_{0}=\frac{1}{2m^{\ast}}[-\hslash^{2}\frac{\partial^{2}}{\partial x^{2}%
}+(-i\hslash\frac{\partial}{\partial y}+eBx+(eB_{0}/K)\sin Kx)^{2}].\label{1}%
\end{equation}
Since the Hamiltonian does not depend on the $y$ coordinate, the unperturbed
wavefunctions are plane waves in the the $y$-direction. This allows us to
write for the wavefunctions,
\begin{equation}
\phi_{nk_{y}}(\bar{x})=\frac{1}{\sqrt{L_{y}}}e^{ik_{y}y}u_{n}(x),\label{2}%
\end{equation}
with $L_{y}$ being a normalization length in $y$-direction and $\bar{x}$ a 2D
position vector in the $x$-$y$ plane. The hamiltonian in equation (1) can be
expressed as
\begin{align}
H_{0} &  =-\frac{\hslash^{2}}{2m^{\ast}}\frac{\partial^{2}}{\partial x^{2}%
}+\frac{1}{2}m^{\ast}\omega_{c}^{2}(x-x_{0})^{2}\nonumber\\
&  +(\omega_{0}/K)(p_{y}+eBx)\sin Kx+(m^{\ast}\omega_{0}^{2}/4K^{2}%
)[1-\cos2Kx],\label{3}%
\end{align}
where $\omega_{c}=\frac{eB}{m^{\ast}}$ is the cyclotron frequency, $\omega
_{0}=eB_{0}/m^{\ast}$ is the modulation frequency, $x_{0}=-l^{2}k_{y}%
=-\frac{\hslash k_{y}}{m^{\ast}\omega_{c}}$ is the coordinate of cyclotron
orbit center, $l=\sqrt{\frac{\hslash}{m^{\ast}\omega_{c}}}$ is the magnetic
length, and $m^{\ast}$ is the effective mass. We can write the unmodulated
eigenstates in the form $\phi_{nk_{y}}(\bar{x})=\frac{1}{\sqrt{L_{y}}%
}e^{ik_{y}y}u_{n}(x;x_{0}),$ with $u_{n}(x;x_{0})=(\sqrt{\pi}2^{n}%
n!l)^{\frac{-1}{2}}\exp(-\frac{1}{2l^{2}}(x-x_{0})^{2})H_{n}(\frac{x-x_{0}}%
{l}),$ where $u_{n}(x;x_{0})$ is a normalized harmonic oscillator wavefunction
centered at $x_{0}$ and $H_{n}(x)$ are Hermite polynomials with $n$ the Landau
level quantum number\cite{33}. Since we are considering weak modulation
$B_{0}<<$ $B$, we can apply standard perturbation theory to determine the
first order corrections to the unmodulated energy eigenvalues in the presence
of modulation%
\begin{equation}
\varepsilon(n,x_{0})=(n+1/2)\hslash\omega_{c}+G_{n}\cos(\frac{2\pi}{a}%
x_{0})\label{4}%
\end{equation}
where $G_{n}=\hslash\omega_{0}\exp(-u/2)[L_{n}(u)/2+L_{n-1}^{1}(u)],u=\frac
{K^{2}l^{2}}{2}=(\frac{2\pi}{a})^{2}\frac{\hslash}{2m^{\ast}\omega_{c}}$ and
$L_{n}(u),$ $L_{n-1}^{l}(u)$ are Laguerre and associated Laguerre polynomials.
This result has been obtained previously\cite{5,8}. The above equation shows
that the formerly sharp Landau levels are now broadened into minibands by the
modulation potential. Furthermore, the Landau bandwidth ($\sim\mid G_{n}\mid$)
oscillates as a function of $n$, since $L_{n}(u)$ is an oscillatory function
of its index\cite{33}.

Hereafter, we employ the Ehrenreich-Cohen Self-Consistent Field (SCF) approach
to determine the density-density response function\cite{34}. Following the SCF
approach, $\Pi_{0}(\bar{q},\omega)$ is the density-density response function
of the non-interacting electron system, given by
\begin{equation}
\Pi_{0}(\bar{q},\omega)=\frac{1}{A}\underset{n,n^{\prime}}{\sum}%
\underset{k_{y}}{\sum}C_{nn^{\prime}}(\frac{\hslash\bar{q}^{2}}{2m^{\ast
}\omega_{c}})\frac{f(\varepsilon(n^{\prime},k_{y}-q_{y}))-f(\varepsilon
(n,k_{y}))}{\varepsilon(n^{\prime},k_{y}-q_{y})-\varepsilon(n,k_{y}%
)+\hslash\omega+i\hslash\eta},\label{5}%
\end{equation}
where
\[
C_{nn^{\prime}}(x)=\frac{n_{2}!}{n_{1}!}e^{-x}x^{n_{1}-n_{2}}[L_{n_{2}}%
^{n_{1}-n_{2}}(x)]^{2}%
\]
with $n_{1}$= max($n$, $n^{\prime}$), $n_{2}$= min($n$, $n^{\prime}$),
$f(\varepsilon)$ is the Fermi-Dirac distribution function, $\bar{q}$ is the 2D
wave number, $A$ is the area of the system, $x=\frac{\hslash\bar{q}^{2}%
}{2m^{\ast}\omega_{c}}$ and $L_{n}^{l}(x)$ are associated Laguerre
polynomials. The density-density response function of the interacting system
can be expressed as%
\begin{equation}
\Pi(\bar{q},\omega)=\frac{\Pi_{0}(\bar{q},\omega)}{1-v_{c}(\bar{q})\Pi
_{0}(\bar{q},\omega)}\label{6}%
\end{equation}
with $v_{c}(\bar{q})=$ $\frac{2\pi e^{2}}{k\bar{q}}$ is the 2-D Coulomb
potential, $k$ is the background dielectric constant. Using the transformation
$k_{y}\rightarrow-k_{y},$ realizing that $\varepsilon(n,k_{y})$ is an even
function of $k_{y},$ interchanging $n\leftrightarrow n^{\prime}$ we can write
the non-interacting density-density response function given by equation (5) as%
\begin{align}
\Pi_{0}(\bar{q},\omega) &  =\frac{m^{\ast}\omega_{c}}{\pi\hslash a}%
\underset{n,n^{\prime}}{\sum}C_{nn^{\prime}}(\frac{\hslash\bar{q}^{2}%
}{2m^{\ast}\omega_{c}})\overset{a}{\underset{0}{\int}}dx_{0}[f(\varepsilon
(n,x_{0}+x_{0}^{\prime})-f(\varepsilon(n^{\prime},x_{0}))]\nonumber\\
&  \times\lbrack\varepsilon(n,x_{0}+x_{0}^{\prime})-\varepsilon(n^{\prime
},x_{0})+\hslash\omega+i\hslash\eta]^{-1}.\label{7}%
\end{align}
In writing the above equation we converted the $k_{y}$-sum into an integral
over $x_{0}$ and $x_{0}^{\prime}=-\frac{\hslash q_{y}}{m^{\ast}\omega_{c}}$.

\section{Magnetoplasmons modes of MM2DEG and comparison with EM2DEG at zero
temperature}

The plasma modes are obtained from the roots of the longitudinal dispersion
relation from Eqs. (6, 7)%
\begin{equation}
1-v_{c}(\bar{q})\operatorname{Re}\Pi_{0}(\bar{q},\omega)=0\label{8}%
\end{equation}
along with the condition $Im\Pi_{0}(\bar{q},\omega)=0$ to ensure long-lived
excitations. The roots of Eq. (8) give the plasma modes of a MM2DEG as
\begin{equation}
1=\frac{2\pi e^{2}}{\kappa\bar{q}}\frac{2}{\pi l^{2}}\underset{n,n^{\prime}%
}{\sum}C_{nn^{\prime}}\left(  x\right)  \{I_{1}(n,n^{\prime},x_{0}^{\prime
};\omega)+I_{1}(n,n^{\prime},x_{0}^{\prime};-\omega)\},\label{9}%
\end{equation}
with%
\begin{equation}
I_{1}(n,n^{\prime},x_{0}^{\prime};\omega)=P\overset{a}{\underset{0}{\int}%
}dx_{0}\frac{f(\varepsilon(n,x_{0}))}{\varepsilon(n^{\prime},x_{0}%
+x_{0}^{\prime})-\varepsilon(n,x_{0}+x_{0}^{\prime})+\hslash\omega}.\label{10}%
\end{equation}
and P is the principal value. From here on we will only show the dependence of
$I_{1}$ on $\omega$ and suppress the rest such that $I_{1}(n,n^{\prime}%
,x_{0}^{\prime};\omega)\rightarrow I_{1}(\omega).$

SdH and Weiss oscillations are found to occur in the magnetoconductivity of
both electric and magnetically modulated 2DEG. These transport measurements
can be explained without taking into account electron-electron interactions.
In order to investigate collective excitations of the system such as
magnetoplasmons it is essential to consider electron-electron interactions.
Magnetoplasmons arise due to the coherent motion of electrons as a result of
electron-electron interactions. Two types of magnetoplasmons can be
identified: Those arising from electronic transitions involving different
Landau bands (inter Landau band plasmons) and those within a single Landau
band (intra Landau band plasmons). Inter-Landau band plasmons involve the
local 1D magnetoplasma mode and the Bernstein-like plasma
resonances\cite{35,36}, all of which involve excitation energies greater than
the Landau-band separation ($\sim\hslash\omega_{c}$). On the other hand,
intra-Landau band magnetoplasmons resonate at energies comparable to the
bandwidths, and the existence of this new class of modes is due to finite
width of the Landau levels. In a MM2DEG considered here, the Landau bandwidth
($\sim\mid G_{n}\mid$) oscillates as a function of the band index $n,$ since
$L_{n}(u),$ $L_{n}^{l}(u)$ are oscillatory functions of the index $n.$ Such
oscillating bandwidths affect the plasmon spectrum of the intra-Landau band
type, when Landau-band separation is larger than the bandwidth as is the case
considered here, resulting in magnetic Weiss oscillations similar to the
electric Weiss oscillations found in the electrically modulated system. These
oscillations are accompanied by SdH type of oscillatory behavior\cite{29,37}.
Both these oscillations are periodic as a function of inverse magnetic field
($1/B$) but occur with different periods and amplitudes. As we show below,
Weiss oscillations in the magnetoplasmon spectrum of a MM2DEG differ in phase
and amplitude with those of an EM2DEG\cite{29}.

We now examine the inter-Landau-band transitions. These transitions occur
between different Landau bands so we consider $n$ $\neq n^{\prime}$in Eq.(10)
which yields
\[
I_{1}(\omega)=\frac{f(\varepsilon(n))}{(\hslash\omega-\Delta)},
\]
where $\Delta=\left(  \varepsilon(n)-\varepsilon(n^{\prime})\right)  $ with
$\varepsilon(n)=(n+\frac{1}{2})\hslash\omega_{c}$, which permits us to write
the following term in Eq (9) as%
\begin{equation}
(I_{1}(\omega)+I_{1}(-\omega))=2\frac{\Delta f(\varepsilon(n))}{(\hslash
\omega)^{2}-(\Delta)^{2}}.\label{11}%
\end{equation}
Next, we consider the coefficient $C_{nn^{\prime}}(x)$ in Eq.(9) and expand it
to lowest order in its argument (low wave-number expansion). In this case, we
are only considering the $n^{\prime}=n\pm1$ terms. The inter-Landau band
plasmon modes under consideration arise from neighboring Landau bands. Hence
for $n^{\prime}=n+1$ and $x\ll$ 1, using the following associated Laguerre
polynomial expansion\cite{38} $L_{n}^{^{l}}(x)=%
%TCIMACRO{\dsum \limits_{m=0}^{n}}%
%BeginExpansion
{\displaystyle\sum\limits_{m=0}^{n}}
%EndExpansion
(-1)^{m}\frac{(n+l)!}{(l+m)!(n-m)!}\frac{x^{m}}{m!}$ for $l>0$ and retaining
the first term in the expansion for $x\ll$ 1, $C_{nn^{\prime}}(x)$ reduces to
\begin{equation}
C_{n,n+1}(x)\rightarrow(n+1)x,\label{12}%
\end{equation}
and for $n^{\prime}=n-1$ and $x\ll1,$ it reduces to
\begin{equation}
C_{n,n-1}(x)\rightarrow nx.\label{13}%
\end{equation}
Substitution of equations (11) and (12, 13) into equation (9) and replacing
$x=\frac{\hslash\bar{q}^{2}}{2m^{\ast}\omega_{c}}$ yields
\[
1=\frac{2\pi e^{2}}{km^{\ast}}\bar{q}\frac{1}{\omega^{2}-\omega_{c}^{2}}%
(\frac{m\omega_{c}}{\pi\hslash}\underset{n}{\sum}f(\varepsilon(n)).
\]
The term in parenthesis is easily recognized as the unmodulated particle
density $n_{D}=\frac{m\omega_{c}}{\pi\hslash}\underset{n}{\sum}f(\varepsilon
(n)),$ where the summation is over all occupied Landau bands. Defining the
plasma frequency through $\omega_{p,2D}^{2}=\frac{2\pi n_{D}e^{2}}{km}\bar
{q},$ we obtain the inter-Landau-band plasmon dispersion relation
$1=\frac{\omega_{p,2D}^{2}}{\omega^{2}-\omega_{c}^{2}}$ or%
\begin{equation}
\omega^{2}=\omega_{c}^{2}+\omega_{p,2D}^{2},\label{14}%
\end{equation}
here $\omega$ is the well-known local 2D principal plasma frequency. Since we
are interested in the modulation induced effects on the magnetoplasmons in
this system, our focus will be on the intra-Landau band magnetoplasmons as the
dispersion relation for inter-Landau band magnetoplasmons have been discussed
and results displayed in \cite{12,29}.

For the intra-Landau-band excitations spectrum, we need to consider
transitions within a Landau miniband, i.e. $n=n^{\prime}$, $\varepsilon
(n^{\prime})-\varepsilon(n)=0$ and $C_{nn^{\prime}}(x)\rightarrow1$. An
analytical expression of the intra-Landau band plasmon energy $\hslash
\overset{\sim}{\omega}$ can be obtained%
\begin{equation}
\hslash^{2}\overset{\sim}{\omega}^{2}=\frac{16e^{2}}{k\bar{q}\pi}\frac
{m^{\ast}\omega_{c}}{\hslash a}\sin^{2}\left(  \frac{\pi}{a}(x_{0}^{\prime
})\right)  \times A_{n}, \label{15}%
\end{equation}
where%
\[
A_{n}=\underset{n}{\sum}G_{n}\times\int_{0}^{a/2}dx_{0}f(\varepsilon
(n,x_{0}))\cos(Kx_{0}).
\]
At zero temperature ($T=0$),%
\begin{equation}
\hslash^{2}\overset{\sim}{\omega}^{2}=\frac{8e^{2}}{k\bar{q}}\frac{m^{\ast
}\omega_{c}}{\pi\hslash}\sin^{2}\left(  \frac{\pi}{a}(x_{0}^{\prime})\right)
\times\underset{n}{\sum\mid}G_{n}\mid\sqrt{1-\Delta_{n}^{2}}\theta
(1-\Delta_{n}), \label{16}%
\end{equation}
with $\Delta_{n}=\mid\frac{\varepsilon_{F}-\varepsilon(n)}{G_{n}}\mid
,\theta(x)$ the Heaviside unit step function. If we replace the magnetic
modulation term ($G_{n}$) by the electric one, Eq. (16) has the same structure
as Eq. (8) of \cite{29} that pertains to electric modulation. The above
expression for $\hslash\overset{\sim}{\omega}$ has been obtained under the
condition $\hslash\omega>>\mid\varepsilon(n,x_{0}+x_{0}^{\prime}%
)-\varepsilon(n,x_{0})\mid$ as $x_{0}^{\prime}\rightarrow0$ which leads to a
relation between the energy and the Landau level broadening $\hslash
\omega>>\mid2G_{n}\sin(\frac{\pi}{a}x_{0}^{\prime})\sin[(\frac{2\pi}{a}%
)(x_{0}+\frac{x_{0}^{\prime}}{2})]\mid$. This ensures that $Im\Pi_{0}(\bar
{q},\omega)=0$ and the intra-Landau-band magnetoplasmons are undamped. For a
given $G_{n}$, this can be achieved with a small but nonzero $q_{y}$ (recall
that $x_{0}^{\prime}=-\frac{\hslash q_{y}}{m^{\ast}\omega_{c}}$).

In general, the inter- and intra-Landau-band modes are coupled for arbitrary
magnetic field strengths. The general dispersion relation is :
\[
1=\frac{\omega_{p,2D}^{2}}{\omega^{2}-\omega_{c}^{2}}+\frac{\overset{\sim
}{\omega}^{2}}{\omega^{2}}.
\]
This equation yields two modes which are given by
\begin{align*}
\omega_{\pm}^{2}  &  =\frac{1}{2}(\omega_{c}^{2}+\omega_{p,2D}^{2}%
+\overset{\sim}{\omega}^{2})\pm\frac{1}{2}\{(\omega_{c}^{2}+\omega_{p,2D}%
^{2}+\overset{\sim}{\omega}^{2}+2\omega_{c}\overset{\sim}{\omega})\\
&  \times(\omega_{c}^{2}+\omega_{p,2D}^{2}+\overset{\sim}{\omega}^{2}%
-2\omega_{c}\overset{\sim}{\omega})\}^{1/2}%
\end{align*}
which reduces to%
\[
\omega_{+}^{2}=\omega_{c}^{2}+\omega_{p,2D}^{2},
\]
and
\[
\omega_{-}^{2}=\overset{\sim}{\omega}^{2}%
\]
with corrections of order $\overset{\sim}{\omega}^{2}/\omega_{c}^{2}$ and
$\overset{\sim}{\omega}^{2}/\omega_{p,2D}^{2}$. So long as $\mid G_{n}%
\mid<\hslash\omega_{c},$ mixing of the inter-and intra-band modes is small.
Only the intra-Landau-band mode ($\hslash\overset{\sim}{\omega})$ will be
excited in the frequency regime $\hslash\omega_{c}>\hslash\omega\sim$ $\mid
G_{n}\mid.$

The intra-Landau-band plasma energy given by equation (11) is shown
graphically in Fig.(1) as a function of $1/B$. The parameters used are
\cite{5,7,8,26,29,30,31}: $m^{\ast}=0.07m_{e}$, $k=12$, $n_{D}=3.16\times
10^{15}$ m$^{-2}$, $a=380$ nm. We also take $q_{x}=0$ and $q_{y}=0.01k_{F},$
with $k_{F}=(2\pi n_{D})^{1/2}$ being the Fermi wave number of the unmodulated
2DEG in the absence of magnetic field. In numerical evaluation we have taken
the sum over thirty Landau levels which ensures convergence of numerical
results. Numerical evaluation of the dispersion relation was performed in
Mathematica. In Fig(1), modulation induced oscillations in the intra-Landau
band mode are apparent, superimposed on SdH-type oscillations. These
oscillations are periodic as a function of inverse magnetic field ($1/B$). To
gain further insight into the results presented in Fig.(1), we consider
equation (16). In the regime $\hslash\omega_{c}>\left\vert G_{n}\right\vert $,
the unit step function vanishes for all but the highest occupied Landau band,
corresponding ,say, to the band index $N$. The sum over $n$ is trivial and
plasma energy is given as $\hslash\overset{\sim}{\omega}=\mid G_{N}\mid
^{1/2}(1-\Delta_{N}^{2})^{1/4}\theta(1-\Delta_{N})$. The analytic structure
primarily responsible for the SdH type of oscillations is the function
$\theta(1-\Delta_{N})$, which jumps periodically from zero (when the Fermi
level is above the highest occupied Landau band) to unity (when the Fermi
level is contained with in the highest occupied Landau band). On the other
hand, the periodic modulation of the amplitude of the SdH type oscillations is
due to the oscillatory nature of the factor $\mid G_{N}\mid^{1/2}$, which has
been shown to exhibit commensurability oscillations \cite{6,19,21,31}. In the
same figure, we also show the intra-Landau band plasma energy for the
electrically modulated 2DEG\cite{29}. The magnetoplasmons spectrum of the
MM2DEG at a specific Landau level is minimum when the corresponding spectrum
for the EM2DEG is maximum, this confirms that magnetic oscillations in the two
system are out of phase. To have comparable results for the two systems, at
zero temperature, the strength of magnetic modulation potential,
$\hslash\omega_{0},$ has to be $\sim8.5$ times smaller than that of the
electric modulation potential, $V_{0}$. This can be understood if we realize
that this constraint arises due to the step function $\theta(1-\Delta_{N})$
appearing in the dispersion relation at zero temperature. Therefore,
comparable results for the two systems when they are subjected to modulation
of equal strength requires that we carry out a finite temperature calculation
to avoid the constraint imposed by the step function at zero temperature.

\section{Temperature dependent magnetoplasmons mode of MM2DEG and comparison
with EM2DEG}

For the finite temperature calculation of the intra-landau band plasma energy,
we invoke the condition of weak modulation and perform the following expansion
in equation (15)%

\begin{equation}
f(\varepsilon(n,x_{0}))\simeq f(\varepsilon(n))+G_{n}f^{\prime}(\varepsilon
(n))\cos(Kx_{0}), \label{17}%
\end{equation}
where $f^{\prime}(x)=\frac{d}{dx}f(x)$ is the derivative of the Fermi-Dirac
distribution function. With the substitution of this expansion in equation
(15), the integral over $x_{0}$ contains two terms. The integral over $x_{0}$
of the first term vanishes and the integral of the second term yields
intra-Landau band dispersion relation%
\begin{equation}
\hslash^{2}\overset{\sim}{\omega}^{2}=\frac{4e^{2}m^{\ast}\omega_{c}}{k\bar
{q}\pi\hslash}\sin^{2}[\frac{\pi}{a}(x_{0}^{\prime})]\times B_{n}, \label{18}%
\end{equation}
where%
\[
B_{n}=\underset{n}{\sum}G_{n}^{2}\times\lbrack-f^{\prime}(\varepsilon(n))]\ .
\]
To facilitate comparison of the above dispersion relation with the results
obtained for an electrically modulated 2DEG we will determine the asymptotic
expressions of intra-Landau band magnetoplasmon spectrum, where analytic
results in terms of elementary functions can be obtained. Moreover, these
asymptotic expressions will allow us to identify terms responsible for SdH and
Weiss oscillations and how they are affected by temperature.

The asymptotic expression can be obtained by using the following asymptotic
expression for the Laguerre polynomials\cite{5,7,8,26}%
\begin{equation}
\exp^{-u/2}L_{n}(u)\rightarrow\frac{1}{\sqrt{\pi\sqrt{nu}}}\cos(2\sqrt
{nu}-\frac{\pi}{4}). \label{19}%
\end{equation}
Note that the above asymptotic expression for $L_{n}(u)$ is valid for $n\gg1$,
at low magnetic fields when many Landau Levels are filled. We now take the
continuum limit:%
\begin{equation}
n-->\frac{\varepsilon(n)}{\hslash\omega_{c}},\overset{\infty}{\underset{n=0}{%
%TCIMACRO{\dsum }%
%BeginExpansion
{\displaystyle\sum}
%EndExpansion
}}-->%
%TCIMACRO{\dint \limits_{0}^{\infty}}%
%BeginExpansion
{\displaystyle\int\limits_{0}^{\infty}}
%EndExpansion
\frac{d\varepsilon}{\hslash\omega_{c}}. \label{20}%
\end{equation}
In the asymptotic limit, $B_{n}$ that appears in equation (18) can be written
\ as%
\begin{equation}
B_{n}=\frac{\hslash^{2}\omega_{0}^{2}}{\pi\hslash\omega_{c}}\sqrt
{\frac{\hslash\omega_{c}}{u}}\left(  \frac{aK_{F}}{2\pi}\right)  ^{2}%
%TCIMACRO{\dint \limits_{0}^{\infty}}%
%BeginExpansion
{\displaystyle\int\limits_{0}^{\infty}}
%EndExpansion
\frac{d\varepsilon}{\sqrt{\varepsilon}}\frac{\beta g(\varepsilon
)}{[g(\varepsilon)+1)]^{2}}\sin^{2}\left(  2\sqrt{nu}-\frac{\pi}{4}\right)
\label{21}%
\end{equation}
where $g(\varepsilon)=\exp[\beta(\varepsilon-\varepsilon_{F})],\beta=\frac
{1}{K_{B}T}.$

Now assuming that temperature is low such that $\beta^{-1}\ll\varepsilon_{F}$
and replacing $\varepsilon=\varepsilon_{F}+s\beta^{-1}$, we can express the
above integral as%
\[
B_{n}=\frac{\hslash^{2}\omega_{0}^{2}}{\pi\sqrt{u\hslash\omega_{c}%
\varepsilon_{F}}}\left(  \frac{aK_{F}}{2\pi}\right)  ^{2}%
%TCIMACRO{\dint \limits_{0}^{\infty}}%
%BeginExpansion
{\displaystyle\int\limits_{0}^{\infty}}
%EndExpansion
ds\frac{e^{s}}{[e^{s}+1)]^{2}}\sin^{2}\left(  2\sqrt{\frac{u\varepsilon_{F}%
}{\hslash\omega_{c}}}-\frac{\pi}{4}+\sqrt{\frac{u}{\hslash\omega
_{c}\varepsilon_{F}}}s\beta^{-1}\right)
\]
with the result%
\begin{equation}
B_{n}=\frac{\hslash^{2}\omega_{0}^{2}}{2\pi\sqrt{u\hslash\omega_{c}%
\varepsilon_{F}}}\left(  \frac{aK_{F}}{2\pi}\right)  ^{2}\left[  1-A\left(
\frac{T}{T_{a}}\right)  +2A\left(  \frac{T}{T_{a}}\right)  \sin^{2}\left[
2\sqrt{\frac{u\varepsilon_{F}}{\hslash\omega_{c}}}-\frac{\pi}{4}\right]
\right]  \label{22}%
\end{equation}
where $T_{a}$\ is the characteristic damping temperature of Weiss
oscillationgiven by $k_{B}T_{a}=\frac{\hslash\omega_{c}aK_{F}}{4\pi^{2}},$
$\frac{T}{T_{a}}=\frac{4\pi^{2}k_{B}T}{\hslash\omega_{c}aK_{F}}$ and
$A(x)=\frac{x}{\sinh(x)}-^{(x-->\infty)}->=2xe^{-x}.$

From equation(18), the asymptotic expression for intra-Landau band plasmon
spectrum is obtained%
\begin{align}
\hslash^{2}\overset{\sim}{\omega}^{2}  &  =\frac{4\hslash^{2}\omega_{0}%
^{2}e^{2}m^{\ast}\omega_{c}}{k\bar{q}\hslash2\pi^{2}\sqrt{u\hslash\omega
_{c}\varepsilon_{F}}}\left(  \frac{aK_{F}}{2\pi}\right)  ^{2}\sin^{2}\left[
\frac{\pi}{a}(x_{0}^{\prime})\right] \nonumber\\
&  \times\left[  1-A\left(  \frac{T}{T_{a}}\right)  +2A\left(  \frac{T}{T_{a}%
}\right)  \sin^{2}\left(  2\sqrt{\frac{u\varepsilon_{F}}{\hslash\omega_{c}}%
}-\frac{\pi}{4}\right)  \right]  \label{23}%
\end{align}
The above expression is not able to account for the SdH type of oscillations
in the magnetoplasmon spectrum. These oscillations can be accounted for by
expressing the density of states (in the absence of disorder)\cite{5} as
\begin{equation}
D(\varepsilon)=\frac{m^{\ast}}{\pi\hslash}\left(  1-2\cos\left[  \frac
{2\pi\varepsilon}{\hslash\omega_{c}}\right]  \right)  \label{24}%
\end{equation}
and inserting the continuum approximation as $\overset{\infty}{\underset{n=0}{%
%TCIMACRO{\dsum }%
%BeginExpansion
{\displaystyle\sum}
%EndExpansion
}}-->2\pi l^{2}%
%TCIMACRO{\dint \limits_{0}^{\infty}}%
%BeginExpansion
{\displaystyle\int\limits_{0}^{\infty}}
%EndExpansion
D(\varepsilon)d\varepsilon,$ this yields the asymptotic expression for the
intra-Landau band magnetoplasmon dispersion relation for MM2DEG%
\begin{align}
\hslash^{2}\overset{\sim}{\omega}^{2}  &  =\frac{4\hslash^{2}\omega_{0}%
^{2}e^{2}m^{\ast}\omega_{c}}{k\bar{q}\hslash2\pi^{2}\sqrt{u\hslash\omega
_{c}\varepsilon_{F}}}\left(  \frac{aK_{F}}{2\pi}\right)  ^{2}\sin^{2}\left[
\frac{\pi}{a}(x_{0}^{\prime})\right]  \times\{[1-A\left(  \frac{T}{T_{a}%
}\right) \nonumber\\
&  +2A\left(  \frac{T}{T_{a}}\right)  \sin^{2}\left(  2\sqrt{\frac
{u\varepsilon_{F}}{\hslash\omega_{c}}}-\frac{\pi}{4}\right)  ]-4A\left(
\frac{T}{T_{s}}\right)  \cos\left[  \frac{2\pi\varepsilon_{F}}{\hslash
\omega_{c}}\right]  \sin^{2}\left[  2\sqrt{\frac{u\varepsilon_{F}}%
{\hslash\omega_{c}}}-\frac{\pi}{4}\right]  \} \label{25}%
\end{align}
where $\frac{T}{T_{s}}=\frac{2\pi^{2}k_{B}T}{\hslash\omega_{c}},$ $T_{s}$
defines the characteristic damping temperature of the SdH oscillations in the
magnetoplasmon spectrum of MM2DEG.

Following the same approach as discussed above for MM2DEG, we can obtain the
intra-Landau band magnetoplasmon spectrum for EM2DEG%
\begin{equation}
\hslash^{2}\overset{\sim}{\omega}^{2}=\frac{4e^{2}m^{\ast}\omega_{c}}{\hslash
k\bar{q}\pi}\sin^{2}[\frac{\pi}{a}(x_{0}^{\prime})]\times B_{n}, \label{26}%
\end{equation}
where $B_{n}=\sum F_{n}^{2}\times\lbrack-f^{\prime}(\varepsilon(n))]$, and
$F_{n}=V_{0}e^{-u/2}L_{n}(u)$ is the modulation width of the EM2DEG with
$V_{0}$ the amplitude of electric modulation. The corresponding asymptotic
result\ for EM2DEG is%
\begin{align}
\hslash^{2}\overset{\sim}{\omega}^{2}  &  =\frac{4V_{0}^{2}e^{2}m^{\ast}%
\omega_{c}}{k\bar{q}2\pi^{2}\hslash\sqrt{u\hslash\omega_{c}\varepsilon_{F}}%
}\sin^{2}\left(  \frac{\pi}{a}(x_{0}^{\prime})\right)  \times\{[1-A\left(
\frac{T}{T_{a}}\right) \nonumber\\
&  +2A\left(  \frac{T}{T_{a}}\right)  \cos^{2}\left(  2\sqrt{\frac
{u\varepsilon_{F}}{\hslash\omega_{c}}}-\frac{\pi}{4}\right)  ]-4A\left(
\frac{T}{T_{s}}\right)  \cos[\frac{2\pi\varepsilon_{F}}{\hslash\omega_{c}%
}]\cos^{2}[2\sqrt{\frac{u\varepsilon_{F}}{\hslash\omega_{c}}}-\frac{\pi}{4}]\}
\label{27}%
\end{align}
The intra-landau band plasmon dispersion relations obtained for the MM2DEG and
the EM2DEG systems given by equations (25, 27) allow us to identify the terms
responsible for Weiss and SdH oscillations. Moreover, the characteristic
damping temperatures appearing in these expressions carry the effects of
temperature on these oscillations. Comparing equations (25,27), the following
differences can be highlighted:

1) Amplitude of the oscillations (Weiss and SdH) are larger by the factor
$\frac{aK_{F}}{2\pi}$\ in MM2DEG compared to those of EM2DEG.

2) The factor $\sin^{2}\left[  2\sqrt{\frac{u\varepsilon_{F}}{\hslash
\omega_{c}}}-\frac{\pi}{4}\right]  $ that appears in equation (25) for MM2DEG
and the corresponding factor $\cos^{2}\left[  2\sqrt{\frac{u\varepsilon_{F}%
}{\hslash\omega_{c}}}-\frac{\pi}{4}\right]  $ in equation (27) for EM2DEG
results in a $\pi/2$\ phase difference in the oscillations in the
magnetoplasmon spectrum of the two systems

Since equations (25, 27) are the key results of this work, we show the
intra-Landau band magnetoplasmon energy for both the magnetically and
electrically modulated systems as a function of inverse magnetic field in
Fig.(2). The results presented are for equal strength of the two modulations
which is taken to be $V_{0}=\hslash\omega_{0}=1$ meV$.$ The temperature is
$0.4$ K. Rest of the parameters are the same as the case for the zero
temperature results presented in Fig.(1). We observe the modulation induced
effects in the intra-Landau band mode, Weiss oscillations modulating the SdH
oscillations in the magnetoplasmon spectrum. From the figure, we see that the
amplitude of Weiss oscillations in the magnetically modulated system is
greater by a factor of $\sim8.5$ compared to the electrically modulated
system. This can be seen from equations (25, 27) where the difference in the
amplitudes is the factor $\frac{aK_{F}}{2\pi}$ and it is $\sim8.5$ for the
parameters that we have used. Therefore, in the magnetically modulated system
the amplitude is larger by this factor compared to the electrically modulated
system. We also observe that Weiss oscillations in the MM2DEG are out of phase
by $\pi/2$ compared to those in EM2DEG. To see the effects of temperature on
Weiss and SdH oscillations in a MM2DEG, we plot the intra-Landau band plasmon
energy for a MM2DEG as a function of inverse magnetic field at two different
temperatures in Fig.(3). The modulation strength is $1$ meV$.$ The results are
shown at the following two temperatures: $0.3$ K and $3$ K$.$ The SdH
oscillations are completely damped at $3$ K whereas Weiss oscillations persist
at this temperature. Eqs.(25, 27) also allow us to determine the temperature
scales for damping of Weiss and SdH oscillations in the magnetoplasmon
spectrum. For a MM2DEG, from Eq.(25), $\frac{T_{a}}{T_{s}}=\frac{aK_{F}}{2}%
\gg1$; e.g., $n_{D}=3.16\times10^{15}$m$^{-2}$ and $a=380$ nm, we have
$\frac{T_{a}}{T_{s}}=27$ for the experimentally relevant parameters considered
here. Hence, SdH oscillations are completely damped at a much lower
temperature compared to Weiss oscillations. These results are consistent with
and complement those obtained from electron transport studies of a
magnetically modulated 2DEG \cite{5,6,7,8,19,20,21,26,27}.

\section{Magnetoplasmon spectrum with Periodic Electric and Magnetic
Modulation: In-phase}

In this section, we calculate the magnetoplasmon spectrum when electric and
magnetic modulations are in-phase. We take the magnetic modulation to have the
same phase as given in the previous section with the in-phase electric
modulation. The energy eigenvalues are\cite{5,26}%
\begin{equation}
\varepsilon(n,x_{0})=(n+1/2)\hbar\omega_{c}+(G_{n}+F_{n})\cos(Kx_{o})
\label{28}%
\end{equation}
and the bandwidth can be written as%
\begin{equation}
\Delta(in-phase)=\frac{2\hslash\omega_{0}ak_{F}\times\sqrt{1+\delta^{2}}}%
{2\pi\sqrt{\pi\sqrt{nu}}}\times\sin\left(  2\sqrt{nu}-\frac{\pi}{4}%
+\Phi\right)  \label{29}%
\end{equation}
where the ratio between the two modulation strengths $\delta=\frac{2\pi V_{0}%
}{\hslash\omega_{0}ak_{F}}=\tan(\Phi).$ The flat band condition from the above
equation is $2\sqrt{nu}-\frac{\pi}{4}+\Phi=i\pi$ where $i$ is an integer. This
condition can also be expressed as $\frac{\sqrt{2n}}{a}l=i+\frac{1}{4}%
-\frac{\Phi}{\pi}$, where $n=n_{F}=\frac{\varepsilon_{F}}{\hbar\omega_{c}%
}-\frac{1}{2}$ is the highest Fermi integer. We see that the flat band
condition in this case depends on the relative strength of the two modulations.

Following the same approach as discussed in the previous section for the
MM2DEG, we can obtain the intra-Landau band magnetoplasmon spectrum in the
presence of in phase modulations as%
\begin{equation}
\hslash^{2}\overset{\sim}{\omega}^{2}=\frac{4e^{2}m^{\ast}\omega_{c}}{k\bar
{q}\pi\hslash}\sin^{2}[\frac{\pi}{a}(x_{0}^{\prime})]\times I_{n}, \label{30}%
\end{equation}
where%
\[
I_{n}=\underset{n}{\sum}(G_{n}+F_{n})^{2}\times\lbrack-f^{\prime}%
(\varepsilon(n))]\ .
\]
In Fig.(4) we show the in-phase magnetoplasmon spectrum (the magnetic and the
electric modulations are in-phase) $\hslash\overset{\sim}{\omega}$ given by
Eq.(30) as a function of the inverse magnetic field for temperature $T=0.3$ K,
electron density $n_{e}=3\times10^{11}$ cm$^{-2},$ the period of modulation
$a=380$ nm$.$ The strength of the electric modulation $V_{0}=0.2$ meV whereas
$B_{0}=0.02$ T which corresponds to $\hslash\omega_{0}=0.03$ meV. In the same
figure we have also shown the magnetoplasmon spectrum when either the magnetic
or electric modulation alone is present. The $\frac{\pi}{2}$ phase difference
in the bandwidths results in the same phase difference appearing in the
magnetoplason spectrum for electric and magnetic modulations as can be seen in
the figure. To better understand the effects of in-phase modulations on the
magnetoplasmon spectrum we consider the asymptotic expression of the
magnetoplasmon spectrum given by Eq.(30). The asymptotic expression is given
by%
\begin{align}
\hslash^{2}\overset{\sim}{\omega}^{2}  &  =\frac{2V_{0}^{2}e^{2}m^{\ast}%
\omega_{c}\delta^{-2}}{k\bar{q}2\pi^{2}\hslash\sqrt{u\hslash\omega
_{c}\varepsilon_{F}}}\sin^{2}\left(  \frac{\pi}{a}(x_{0}^{\prime})\right)
\times(1+\delta^{2})\{[1-A\left(  \frac{T}{T_{a}}\right) \nonumber\\
&  +(2A\left(  \frac{T}{T_{a}}\right)  -4A\left(  \frac{T}{T_{s}}\right)
\cos[\frac{2\pi\varepsilon_{F}}{\hslash\omega_{c}}])\sin^{2}[2\sqrt
{\frac{u\varepsilon_{F}}{\hslash\omega_{c}}}-\frac{\pi}{4}+\Phi]\} \label{31}%
\end{align}
From the asymptotic expression given by Eq.(31), we observe that in the
presence of in-phase electric and magnetic modulatio the magnetoplasmon energy
acquires a dependence on the phase factor $\Phi$ and $\delta$ which depend on
the relative modulation strengths. The shift in the Weiss oscillations when
in-phase electric and magnetic modulations are present can be seen in Fig.(5).
How the Weiss oscillations are affected as $\Phi$ as well as the magnetic
field is varied can be seen in Fig.(5). The results shown are for a fixed
magnetic modulation of strength $\hslash\omega_{0}=0.03$ meV and the electric
modulation is varied. The change in $V_{0}$ results in a corresponding change
in both $\delta$ and $\Phi$. From Fig.(5), we observe that the position of the
extrema in the magnetoplasmon spectrum as a function of the inverse magnetic
field depend on the relative strength of the modulations.

The effects of electric and magnetic modulations that are out-of-phase on the
magnetoplasmon spectrum can be better appreciated if we consider the
asymptotic expression in this case. This is taken up in the next section.

\section{Magnetoplasmons with Periodic Electric and Magnetic Modulation:
Out-of-phase}

In this section, we calculate the magnetoplasmon spectrum when electric and
magnetic modulations are out of phase by $\pi/2.$ We consider magnetic
modulation out of phase with the electric one: We take the electric modulation
to have the same phase as given in the previous section with the $\pi/2$ phase
difference incorporated in the magnetic field. The energy eigenvalues
are\cite{5,26}%
\begin{equation}
\varepsilon(n,x_{0})=(n+1/2)\hbar\omega_{c}+\sin(Kx_{0})G_{n}+F_{n}\cos
(Kx_{0}), \label{32}%
\end{equation}
and the bandwidth is%
\begin{equation}
\Delta(\text{out of phase})=\frac{2\hslash\omega_{0}ak_{F}}{2\pi\sqrt{\pi
\sqrt{nu}}}\times\sqrt{\delta^{2}+(1-\delta^{2})\sin\left(  2\sqrt{nu}%
-\frac{\pi}{4}\right)  }\,. \label{33}%
\end{equation}
The term responsible for Weiss oscillations is the $\sin\left(  2\sqrt
{nu}-\frac{\pi}{4}\right)  $ term under the square root which can be readily
seen by considering the large $n$ limit of the bandwidth. Therefore for
$\delta=\pm1$ Weiss oscillations are no longer present in the bandwidth.

Following the same approach discussed in the previous section for MM2DEG, we
can obtain the intra-Landau band magnetoplasmon spectrum in the presence of
out of phase modulations as
\begin{equation}
\hslash^{2}\overset{\sim}{\omega}^{2}=\frac{4e^{2}m^{\ast}\omega_{c}}{k\bar
{q}\pi\hslash}\sin^{2}[\frac{\pi}{a}(x_{0}^{\prime})]\times O_{n}, \label{34}%
\end{equation}
where%

\[
O_{n}=\underset{n}{\sum}(G_{n}^{2}+F_{n}^{2})\times\lbrack-f^{\prime
}(\varepsilon(n))]\ .
\]

The asymptotic expression in the presence of both electric and magnetic
modulations that are out of phase is obtained by substituting the asymptotic
expressions for the Laguerre polynomials and converting the sum into
integration with the result
\begin{align}
\hslash^{2}\overset{\sim}{\omega}^{2}  &  =\frac{2V_{0}^{2}e^{2}m^{\ast}%
\omega_{c}\delta^{-2}}{k\bar{q}2\pi^{2}\hslash\sqrt{u\hslash\omega
_{c}\varepsilon_{F}}}\sin^{2}\left(  \frac{\pi}{a}(x_{0}^{\prime})\right)
\times\{2\delta^{2}+(1-\delta^{2})[1-A\left(  \frac{T}{T_{a}}\right)
\nonumber\\
&  +2A\left(  \frac{T}{T_{a}}\right)  \sin^{2}[2\sqrt{\frac{u\varepsilon_{F}%
}{\hslash\omega_{c}}}-\frac{\pi}{4}]]-4A\left(  \frac{T}{T_{s}}\right)
\cos[\frac{2\pi\varepsilon_{F}}{\hslash\omega_{c}}])[\delta^{2}-(\delta
^{2}-1)\sin^{2}[2\sqrt{\frac{u\varepsilon_{F}}{\hslash\omega_{c}}}-\frac{\pi
}{4}]]\} \label{35}%
\end{align}
From the expression of the out-of-phase bandwidth given by Eq.(33) we find
that Weiss oscillations in the bandwidth are absent for relative modulation
strength $\delta=\pm1$, the same is reflected in the magnetoplasmon spectrum
as the term responsible for Weiss oscillations ($\sin^{2}[2\sqrt
{\frac{u\varepsilon_{F}}{\hslash\omega_{c}}}-\frac{\pi}{4}]$) vanishes for
$\delta=\pm1$ as can be seen from the above equation. Therefore the
magnetoplasmon spectrum does not exhibit Weiss oscillations when the relative
modulation strength $\delta=\pm1.$The spectrum as a function of magnetic field
when the electric and magnetic modulations are out-of-phase is shown in
Fig.(6). The results shown are for a fixed magnetic modulation of strength
$\hslash\omega_{0}=0.03$ meV and the electric modulation $V_{0}$ is allowed to
vary between positive and negative values. The other parameters are the same
as in Figs.(4, 5). As $V_{0}$ is varied there is a corresponding change in
$\delta.$ We find that the positions of the extrema of the spectrum as a
function of the inverse magnetic field do not change as $\delta$ is varied
since the phase factor $\Phi$ does not appear in the expression of
magnetoplasmon spectrum when the two modulations are out of phase. It is also
observed in Fig.(6) that there is a $\frac{\pi}{2}$ phase difference between
the curves for $\delta\geq1$ and $\delta<1.$ The same behavior is observed in
the bandwidth which is reflected in the magnetoplasmon spectrum.

\section{Conclusions}

In conclusion, we have determined the inter and intra-Landau band
magnetoplasmon spectrum for a magnetically modulated two-dimensional electron
gas in the presence of a perpendicular magnetic field. Our results show that
magnetic Weiss oscillations occur in intra-Landau band magnetoplasmon
spectrum. Their origin lies in the interplay of the two physical length scales
of the system i.e. the modulation period, and cyclotron diameter at the Fermi
level. When the strength of magnetic modulation potential is equal to the
electric one, the magnetic Weiss oscillations in the magnetoplasmon spectrum
are out of phase and occur with a larger amplitude compared to the electric
Weiss oscillations in a 2DEG. These results also exhibit that the Weiss
oscillations depend on the temperature much less than that of the SdH
oscillations. We have determined the effects of both the electric and magnetic
modulations on the magnetoplasmon spectrum of 2DEG. These oscillations are
affected by the relative phase of the two modulations and positions of the
extrema of the oscillations depend on the relative strength of the two
modulations. We find complete suppression of Weiss oscillations for particular
relative strength of the modulations when the modulations are out-of-phase.

K. Sabeeh would like to acknowledge the support of the Pakistan Science
Foundation (PSF) through project No. C-QU/Phys (129). M. Tahir would like to
acknowledge the support of the Pakistan Higher Education Commission (HEC).

*Electronic address: m.tahir06@imperial.ac.uk.

$^{\dag}$Electronic address: ksabeeh@qau.edu.pk, kashifsabeeh@hotmail.com

\end{document}